# The Geometric Mean Squared Displacement and the Stokes-Einstein Scaling in a Supercooled Liquid.

Shibu Saw and Peter Harrowell

School of Chemistry, University of Sydney, Sydney NSW 2006 Australia

Abstract

It is proposed that the rate of relaxation in a liquid is better described by the geometric mean of the van Hove distribution function, rather than the standard arithmetic mean used to obtain the mean squared displacement. The difference between the two means is shown to increase significantly with an increase in the non-Gaussian character of the displacement distribution. Preliminary results indicate that the geometric diffusion constant results in a substantial reduction of the deviation from Stokes-Einstein scaling.

PACS: 66.20.Cy, 64.70.Q-

There is a long standing interest in liquid theory in the possibility of relating the relaxation of stress or structure in a liquid to some statistic of the time dependent distribution of particle displacements, the van Hove distribution function [1]. The failure of the mean squared displacement to achieve this in supercooled liquids has been well documented in the form of the 'breakdown' of the Stokes-Einstein relation between the diffusion constant and the viscosity [2]. In this note we argue that the quantity of interest in these relaxation processes is the number of 'distinct' configurations sampled and that this number is directly related to the geometric mean of the displacement distribution.

Relaxation in a dense liquid is a consequence of the system exploring a sufficient number of statistically distinct configurations so as to de-correlate from the initial configuration. As the number of distinct configurations should be proportional to the volume of the configuration space explored over a given time interval, it is the growth of this volume that should best correlate with the relaxation dynamics.

Consider a liquid comprised on N spherically symmetric particles with no internal degrees of freedom. The configuration space of a d-dimensional liquid is defined as the space of $\chi = d \times N$ dimensions corresponding to the position vectors {r} of the particles. Let the particle displacements, after a time interval t, be

$$\Delta \vec{r}_i(t) = \vec{r}_i(t) - \vec{r}_i(0) \tag{1}$$

with $\Delta \vec{r}_i(t) = \{x_{\alpha,i}(t)\}$ where $x_{\alpha,i}$ is the value of a one Cartesian component of the displacement vector of the ith particle. We define a dxN dimensional volume $\Omega(t)$ corresponding to an enclosing envelope of the configuration space explored by the set of N such displacement vectors as

$$\Omega(t) = \prod_{k=1}^{\chi} |x_k(t)| \tag{2}$$

The relation between this volume and the true volume of the sampled configuration space is non-trivial since the exclusion of particle overlap renders part of $\Omega$ inaccessible. A case can be made, however, that $\Omega(t)$ is proportional to the number of distinct states (e.g. local potential energy minima) sampled in the time interval $t$. The argument is adapted from Frenkel et al [3]. Let <υ> be the average dN-dimensional volume of a single local potential energy minimum. The number χ(t) of these minima in the volume Ω(t) is $\Omega(t)/<v>$. As long as we can approximate <υ> as being independent of $\Omega(t)$, the proportionality between χ(t) and $\Omega(t)$ should hold.

$\Omega(t)$ is directly related to the *geometric* mean of the the van Hove distribution. The mean squared displacement associated with same particle motions is, in contrast, an *arithmetic* mean, i.e.

$$<\Delta r^2(t)> = \frac{1}{\chi}\sum_k^\chi x_k^2(t) \qquad (3)$$

As proven by Cauchy [4], the arithmetic mean of a distribution is always greater than or equal to the geometric mean, with equality only possible if all members of the distribution are equal, i.e. the distribution is a delta function. The actual situation in liquids is far from this homogeneous limit. In the context of alternate averages of the particle displacements, we note that Flenner and Szamel [5] introduced a measure of non-Gaussian behaviour (see below) based on the ratio of the arithmetic and *harmonic* mean squared displacements, where the latter is given by $\left\langle \frac{1}{\Delta r^2} \right\rangle^{-1}$.

If we were to assume that each Cartesian component of the particle displacements followed a Gaussian distribution with the same variance, i.e.

$$P(x,t) = \frac{2}{\sigma(t)\sqrt{\pi}}\exp\left(-\frac{x^2}{\sigma^2(t)}\right), \qquad (4)$$

then we find that

$$\Omega^{1/\chi}(t) = \sigma(t) \times \frac{\exp(-C/2)}{2} \approx \sigma(t) \times 0.375 \qquad (5)$$

where C is Euler's constant, and

$$<\Delta r^2(t)>^{1/2} = \sigma(t) \times \frac{1}{\sqrt{2}} \qquad (6)$$

Comparing Eqs. 5 and 6 we find that the two averages are linearly related. This means that when the displacements can be described by a single Gaussian distribution, then the volume of configuration space can be simply related to the mean squared displacement by

$$\Omega(t) \propto <\Delta r^2(t)>^{\chi/2} \qquad (7)$$

A simple Gaussian distribution of displacements, however, is valid in liquids only as a long time limit. The failure of this approximation in supercooled liquids has been the subject of considerable study [4]. Even in the equilibrium liquid, non-Gaussian behaviour is observed over the time interval corresponding to that between 3 and 10 collisions [5]. One measure of this deviation is the non-Gaussian parameter,

$$\alpha(t) = \frac{<x^4>}{3<x^2>^2} - 1 \qquad (8)$$

which is found [4] to rise from zero at short times to reach a maximum at a time roughly that for structural relaxation before decaying back to zero for longer times. This time dependence of the non-Gaussian character of low temperature dynamics has been explained as follows. Over short times the particle motions are too small to resolve local variations in mobility while at very long times, the mixing of different dynamic environments results in a homogeneous dynamics. It is only in the intermediate time scales that the dynamic heterogeneities, associated with the intermittent trapping of some particles, is resolved.

A simple prescription for a non-Gaussian displacement distribution P(x) is to use a sum of *M* Gaussians with different variances, i.e.

$$P(x) = \sum_{m=1}^{M} c_M(t) \frac{2}{\sigma_m(t)\sqrt{\pi}} \exp\left(-\frac{x^2}{\sigma_m^2(t)}\right) \tag{9}$$

where $\sum_{m=1}^{M} c_m(t) = 1$ and

$$\alpha(t) = \frac{\sum_{m=1}^{M} c_m(t)\sigma_m^4(t)}{\left(\sum_{m=1}^{M} c_m(t)\sigma_m^2(t)\right)^2} - 1 \tag{10}$$

The expressions for $\Omega(t)$ and $<\Delta r^2(t)>$ are now

$$\Omega^{1/\chi}(t) = \frac{\exp(-C/2)}{2} \prod_{m=1}^{M} \sigma_m^{c_m(t)}(t) \tag{11}$$

$$<\Delta r^2(t)>^{1/2} = \left(\sum_{m=1}^{M} c_m(t) \frac{\sigma_m^2(t)}{2}\right)^{1/2} \tag{12}$$

To demonstrate the influence of non-Gaussian statistics, it is sufficient to consider a sum of just two Gaussians in Eq.9. We shall set $\sigma_1 = 1.0$ and $c_1 = 1 - c_2$, by construction, so that leaves us with the parameter space $(\sigma_2, c_2)$. In Fig. 1 we plot the ratio of $\Omega^{1/\chi}$ over $<\Delta r^2>^{1/2}$ as a function of the non-Gaussian parameter α.

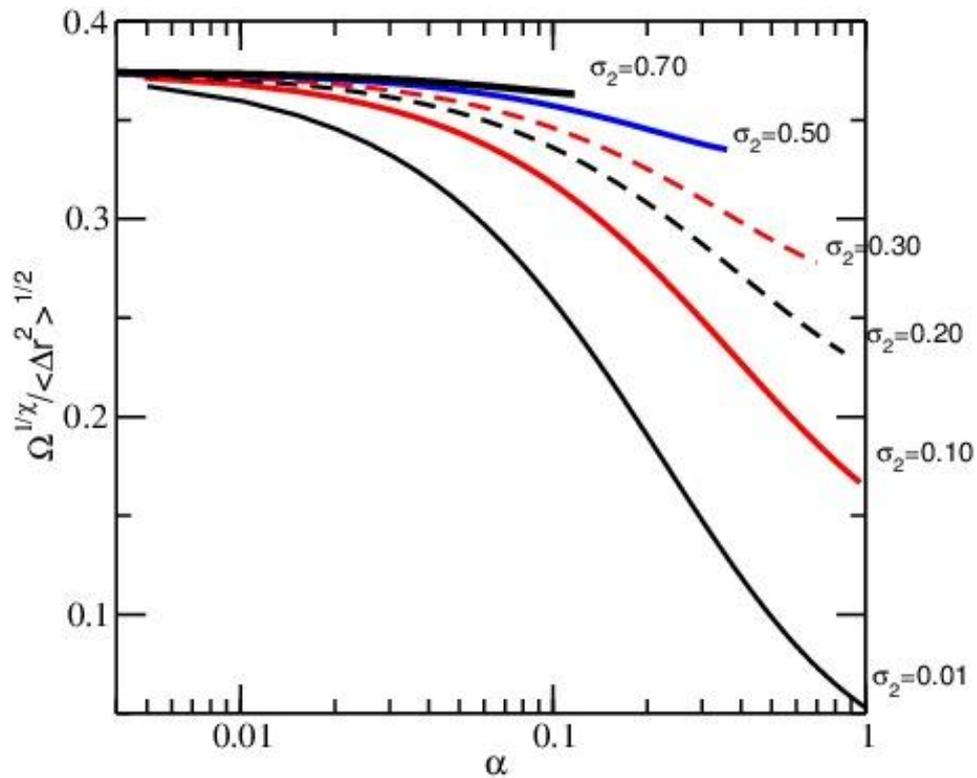

**Figure 1.** The ratio $\dfrac{\Omega^{1/\chi}}{<\Delta r^2>^{1/2}}$ plotted against the non-Gaussian parameter α for the model distribution P(x) (Eq. 9). Each curve corresponds to a fixed value of $\sigma_2$ (values indicated), and the variation of $c_2$ from 0 (left hand side) to 0.5 (right hand side).

The results in Fig. 1 clearly demonstrate that, as the deviation from Gaussianity increases, so to does the discrepancy between the volume of configuration space sampled and the estimation of this quantity using the mean squared displacement. This result is consistent with the common observation [2,6] that relaxation rates in a supercooled liquid cease to scale with the diffusion constant as the non-Gaussian parameter increases with increasing supercooling. While increasing the non-Gaussian parameter does result in a decrease in $\dfrac{\Omega^{1/\chi}}{<\Delta r^2>^{1/2}}$, the data in Fig. 1 makes clear that the magnitude of this trend is sensitive to other details of the distribution. This means that the failure of $<\Delta r^2(t)>$ to provide a useful predictor of relaxation rates in supercooled liquids can only be partially resolved, at best, by the correction provided by the non-Gaussian parameter.

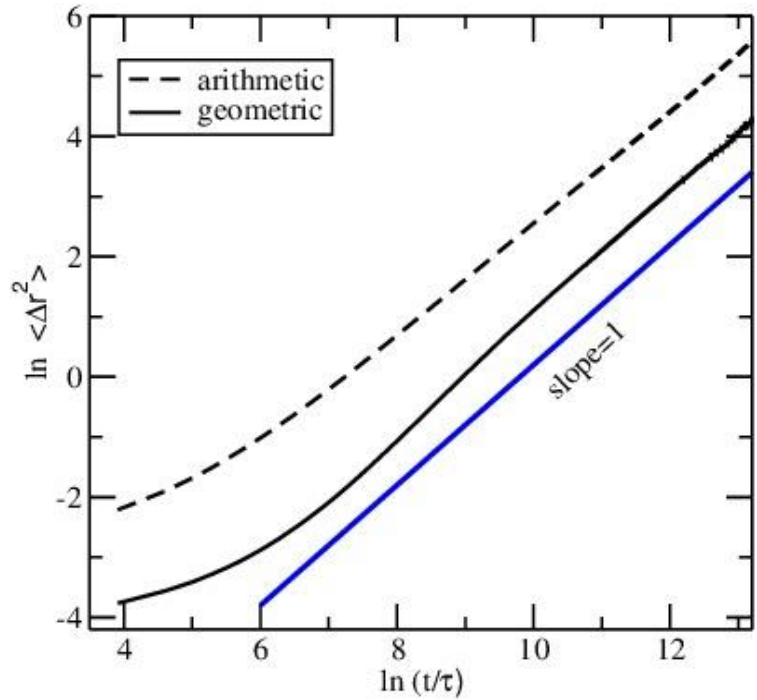

**Figure 2.** The log-log plot of the mean squared displacement $<\Delta r^2>$ vs time of the A particles for the binary mixture at T = 0.415 calculated as the arithmetic and geometric mean. Both curves reach an asymptotic slope of one, consistent with Fickian diffusion, but with different diffusion constants.

As already suggested by Eq. 7, the long time limits of the arithmetic and geometric $<\Delta r^2>$ increase linearly in time as shown in Fig. 2. The results are from molecular dynamics simulations of an equimolar mixtures of soft disks whose parameters and simulation details can be found in ref. 8. The associated diffusion constant can be extracted as

$D = \lim_{t \to \infty} \dfrac{d<\Delta r^2>}{dt}$. The Stokes-Einstein (SE) scaling [2] refers to the constancy of the ratio

ηD/T, where η is the shear viscosity. In Fig. 3 we plot this quantity, normalized so that the high temperature value is set to one, for each choice of the diffusion coefficient. The breakdown is SE scaling is evident for the points associated with the arithmetic diffusion constant. Even given the poor statistics for both transport coefficients at low T, we find that the use of the geometric diffusion constant significantly reduces the deviation from the SE scaling at low temperature.

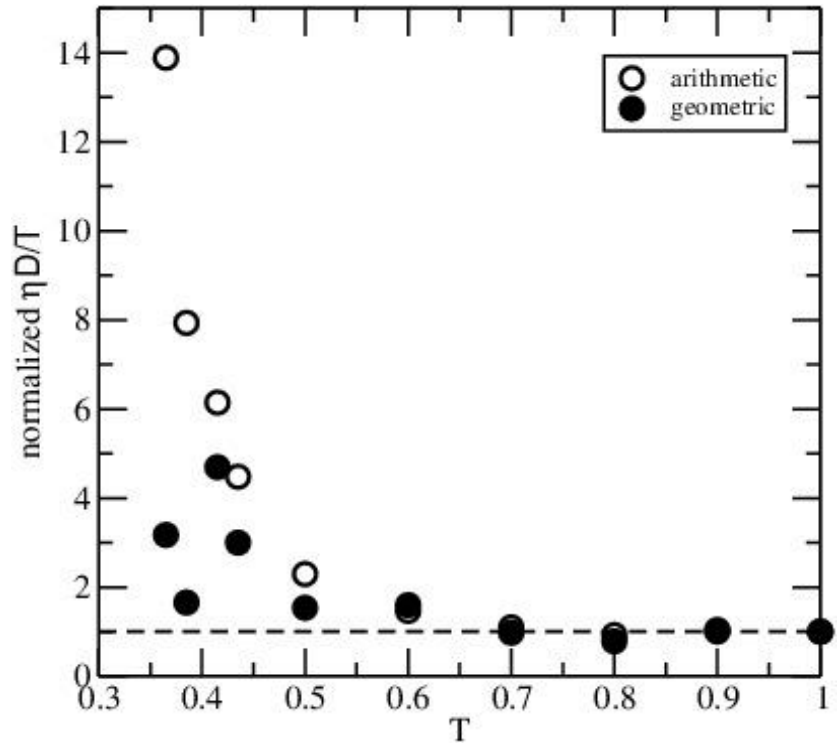

**Figure 3**. The quantity $\eta D/T$ is plotted against T using the diffusion constants for the A particles obtained by arithmetic and geometric averaging, as indicated. To clarify the high temperature Stokes-Einstein behaviour, $\eta D/T$ has been normalized in each case so that the high T limit is set to one with a horizontal dashed line included as a guide to the eye.

In conclusion, since relaxation is a direct reflection of the number of distinct configurations that have been accessed within a given time, we propose that the volume Ω(t), defined in Eq.

2, is the statistic of the particle motions to use as a measure of relaxation time scales. The *arithmetic* mean squared displacement is not as useful in the analysis of relaxation, irrespective of its central role in evaluating the diffusion constant, due to the weight it gives large magnitude outliers of the distribution. We have presented preliminary results demonstrating that a geometric diffusion constant can be consistently obtained and, when used in the quantity $\eta D/T$, maintains the SE scaling to lower temperatures than is achieved using the arithmetic diffusion constant. We believe that these results will provide a useful new approach to understanding slow relaxation in terms of the underlying particle motions responsible.

**Acknowledgements**

This work was supported by a Discovery grant from the Australian Research Council.